\documentclass{emulateapj}

\usepackage{graphicx}
\usepackage{natbib}
\slugcomment{ApJ Accepted, Jan. 23 2008}
\shorttitle{Infrared Measurement of Quasar Obscuration}
\shortauthors{Treister, Krolik and Dullemond}

\begin{document}
\title{Measuring the Fraction of Obscured Quasars by the Infrared Luminosity
of Unobscured Quasars}

\author{Ezequiel Treister,\altaffilmark{1} Julian H. Krolik,\altaffilmark{2} and
Cornelis Dullemond\altaffilmark{3}}
\altaffiltext{1}{European Southern Observatory, Casilla 19001, Santiago 19, Chile. Email: etreiste@eso.org}
\altaffiltext{2}{Department of Physics and Astronomy, Johns Hopkins University, 3400 North Charles Street, Baltimore, MD 21218-2686.Email: jhk@pha.jhu.edu}
\altaffiltext{3}{Max-Planck-Institut f\"ur Astronomie, K\"onigstuhl 17, 69117 Heidelberg,
Germany. Email: dullemon@mpia.de}

\begin{abstract} 
Recent work has suggested that the fraction of obscured AGN declines with
increasing luminosity, but it has been difficult to quantify this trend. Here, we
attempt to measure this fraction as a function of luminosity by studying the
ratio of mid-infrared to intrinsic nuclear bolometric luminosity in unobscured
AGN.  Because the mid-infrared is created by dust reprocessing of shorter wavelength
nuclear light, this ratio is a diagnostic of $f_{\rm obsc}$, the fraction of solid angle
around the nucleus covered by obscuring matter. In order to eliminate possible
redshift-dependences while also achieving a large dynamic range in luminosity, we have
collected archival 24 micron MIPS photometry from objects with $z$$\sim$1 in the
Sloan Digital Sky Survey (SDSS), the Great Observatories Origins Deep Survey (GOODS) and
the Cosmic Evolution Survey (COSMOS). To measure the bolometric luminosity for each
object, we used archival optical data supplemented by GALEX data. We
find that the mean ratio of 24~$\mu$m to bolometric luminosity decreases by a factor of
$\sim$3 in the $L_{\rm bol}$=10$^{44}$-3$\times$10$^{47}$~ergs~s$^{-1}$ range, but there
is also a large scatter at constant $L_{\rm bol}$. Using radiation transfer
solutions for model geometries, we show how the IR/bolometric ratio relates to
$f_{\rm obsc}$ and compare these values with those obtained obtained from samples of
X-ray selected AGN.  Although we find approximate agreement, our method indicates
somewhat higher values of $f_{\rm obsc}$, particularly in the middle range of luminosities,
suggesting that there may be a significant number of heavily obscured AGN
missed by X-ray surveys.
\end{abstract}


\section{Unified AGN Schemes at Low Luminosity}

Following the initial recognition by \citet{antonucci85} that the nucleus of the
prototypical type~2 Seyfert galaxy NGC~1068 must be surrounded by a toroidal belt of gas
and dust, a tremendous amount of evidence has accumulated in support of the idea that
similar obscuring tori surround the nuclei of other type~2 Seyfert galaxies (as reviewed,
for example, in \citealp{antonucci93,krolik99}).  Spectropolarimetry (the method
originally employed by Antonucci and Miller) applied to both type~2 Seyfert galaxies and
radio galaxies \citep{diserego94} often reveals in polarized light broad optical/UV
emission lines and a strong non-stellar continuum that are essentially invisible in the
total flux spectrum.  In many examples of both these types of AGN, conical regions of
bright line emission from highly-ionized elements point at the otherwise obscured nucleus
\citep{ferruit00,schmitt03}.  Similarly, both type~2 Seyfert galaxies and radio galaxies
often exhibit X-ray continua that either show evidence for very large column densities of
material along the line of sight or are extremely weak, likely indicating that the
obscuration is Compton thick \citep{risaliti99,treister04}.  In two cases, NGC~1068 and
the Circinus galaxy, the obscuring torus can be seen directly in interferometric infrared
imaging \citep{jaffe04,tristram07}.

The principal effect of this toroidal obscuration is that observers trying to see the
nucleus along lines of sight that pass through it are prevented from seeing anything but
dust-reprocessed infrared continuum and (perhaps) hard X-rays.  Consequently, most of the
classic signatures of AGN---broad optical/UV emission lines, strong optical/UV non-stellar
continuum, strong X-ray continuum---are obliterated when the object is seen from such a
direction.  Only when the line of sight passes through the central hole of the torus can
all these features be seen, and the object is perceived as a type~1 AGN.

In the nearby Universe, where we can see large numbers of low luminosity AGN, large
statistical samples have been assembled in order to measure the luminosity functions of
both type~1 and type~2 AGN. Ideally compared at matched bolometric luminosity, but more
often in terms of their luminosity in a single band or feature, the ratio of their numbers
can be immediately interpreted as the ratio of unobscured to obscured solid angle.  For
example, \citet{hao05} and \citet{simpson05}, using samples of optically-selected AGN from
the Sloan Digital Sky Survey, found that the ratio of type~1 objects to type~2
objects increases with increasing luminosity,
from $\simeq 2/3$ at an [OIII]~5007 luminosity of $10^6L_{\odot}$ to $\simeq 2$ for line
luminosities $\sim 3 \times 10^7L_{\odot}$. Using samples of X-ray selected AGN,
\citet{steffen03} and \citet{barger05} argued for a similar shift in the ratio of
unobscured to obscured, suggesting that the unobscured variety come to dominate the total
population for 2--8~keV luminosity $\gtrsim 10^{44}$~erg~s$^{-1}$. Similar results were
obtained by \citet{ueda03}, \citet{lafranca05} and others, also using X-ray selected AGN
samples. In fact, hints of such a trend were already seen in the much smaller 3CR radio
sample \cite{lawrence91}.

However, there remain significant uncertainties in the measured population ratio at high
luminosity.  For example, Compton-thick AGN are completely missing from X-ray-selected
samples. Other sample construction methods (e.g., the very large sample of type~2 quasars
selected from the SDSS based primarily on [OIII] emission by \citealp{zakamska03}) can
potentially yield quantitative estimates of comparative luminosity functions, but require
substantial work in order to turn catalogs into space densities.  Although much effort has
been made in the last few years to construct AGN samples from mid-infrared observations
with the {\it Spitzer Space Telescope}, the selection effects remain too poorly understood
and the sample size is sometimes too small to permit extraction of a trend in the type
ratio as a function of luminosity \citep{stern05,alonso-herrero06,martinez06,lacy07}.

Here we attempt a different approach to the problem of measuring the unobscured/obscured
ratio: using the ratio of reradiated mid-infrared continuum to bolometric luminosity in
type~1 AGN as a surrogate (cf. the suggestions in \citealp{lawrence91,barger05}).  Because
the obscuration transforms essentially all the nuclear radiation incident upon it into
mid-infrared continuum via dust reradiation, and detailed radiation transfer calculations
suggest that most of that continuum is radiated toward
observers in the type~1 direction no matter what the detailed arrangement of dust may be
\citep{pier92,granato94,efstathiou95,nenkova02,vanbemmel03,dullemond05}, this ratio should
give a good indicator of the ratio of unobscured/obscured solid angle.  The relative ease
of identifying type~1 AGN across a wide dynamic range in luminosity also gives this method
a number of advantages relative to those dependent on actually counting type~2 objects.
\citet{maiolino07} have recently made an effort to implement this program, but
the details of their method differ in significant ways from ours; we will comment on
specific contrasts as they arise.

This paper is structured as follows: In section \S2 we present the selection criteria and
basic observational properties of the sources used in this work. In section \S3 we study
the correlations found in this sample, while the significance and interpretation of these
trends are discussed in \S 4. Our conclusions are presented in \S 5. When required, we
assume a $\Lambda$CDM cosmology with $h_0$=0.71, $\Omega_m$=0.3 and $\Omega_\Lambda$=0.7,
in agreement with the most recent cosmological observations \citep{spergel07}.

\section{Sample Definition and Data Sources}

In order to eliminate any possibility that what we find involves an evolutionary effect
rather than a luminosity dependence, we chose a sample of AGN that all have nearly the
same redshift ($0.8 \leq z \leq 1.2$), in contrast with the \citet{maiolino07} sample
which includes sources at all redshifts. This particular redshift is a convenient choice
because it is large enough that luminous quasars were common, but not so large that low
luminosity AGN are too faint to detect. To achieve the greatest possible dynamic range in
luminosity, we made use of three samples with complementary properties: one that has very
large solid angle but is relatively shallow (Sloan Digital Sky Survey: SDSS), one that has
smaller solid angle but goes deeper (Great Observatories Origins Deep Survey: GOODS) and
one intermediate approach (Cosmic Evolution Survey: COSMOS).

For our infrared band, we used the Spitzer Space Telescope MIPS $24\mu$m channel.  The
corresponding rest-frame wavelengths are $\simeq$12$\mu$m; by requiring $z$$\leq$1.2, we
avoid most of the contamination from the $9.7\mu$m silicate feature. For the highest
redshift sources, this silicate feature can influence our 24$\mu$m photometry, however
this will affect only a small fraction of our sample. In addition, at this relatively
short wavelength, stellar-heated dust is usually relatively faint. From the composite QSO
spectrum with individually detected PAH features of Schweitzer et al. (2006; Figure 2 top
panel), we can estimate the fraction of integrated flux contributed by the 11.3 $\mu$m PAH
feature in the MIPS 24 micron band at 0.8$<$$z$$<$1.2, and we find a maximum contribution
of only $1.3\%$; hence this contribution is completely negligible. Moreover, because we
use only unobscured AGN with $L_{\rm bol}$ from 10$^{44}$ to 10$^{47.5}$ ergs~s$^{-1}$,
stellar heating is unlikely to be a significant contaminant. The AGN from the SDSS were
selected based on their optical properties, while in the case of the GOODS and COSMOS
sources they were X-ray selected. In any case, the selection is completely independent of
their infrared properties, so it will not bias our results.  In only one case, a SDSS
source is located in the field of view of a Spitzer observation, but a counterpart was not
detected. In that case, we estimate the corresponding upper limit as described below.

For our indicator of the nuclear luminosity, we used optical and UV (GALEX) data to
estimate the bolometric luminosity individually for each source.  Because the GALEX NUV
band (1750--2800~\AA~) corresponds to 875--1900~\AA~ in the rest-frame, we are able to
measure directly the flux in much of the spectral peak region.  The details of our method
are presented in \S 2.5.  In addition, we have also found it is sometimes convenient to
use the SDSS i-band luminosity as a standard of comparison because in our redshift range
it corresponds to rest-frame B-band.

\subsection{SDSS Sample Data Collection}

In order to construct a large sample of high luminosity unobscured quasars, we used the
results from the Sloan Digital Sky Survey (SDSS; \citealp{york00}), specifically from data
release 5 \citep{adelman06}\footnote{Data available at http://www.sdss.org/dr5}. Only
sources classified as quasars by the SDSS pipeline, and thus presenting high-ionization
broad emission lines, were considered. A detailed description of the SDSS quasar
classification scheme was presented by \citet{yip04}. There are 11937 SDSS/DR5 quasars in
the 0.8$\leq$$z$$\leq$1.2 redshift range.

Spitzer data for these SDSS quasars were obtained from the archive using the Leopard v6.1
software\footnote{Leopard can be obtained at
http://ssc.spitzer.caltech.edu/propkit/spot/}. A total of 638 Astronomical Observation
Requests (AORs) were found with at least one source in our sample in the field. In many
cases, these correspond to different observations of the same field, so only 206 quasars
in our SDSS/DR5 sample were observed by Spitzer/MIPS. Only the post-BCD (Basic Calibrated
Data) frames for each AOR were downloaded. The Mosaicking and Point-source Extraction
(MOPEX; \citealp{makovoz05}) package version 030106\footnote{MOPEX can be downloaded from
http://ssc.spitzer.caltech.edu/postbcd/download-mopex.html} was used in order to extract
the sources and calculate fluxes from each AOR. The procedure described by
\citet{makovoz05}\footnote{A more detailed description can be found at
http://ssc.spitzer.caltech.edu/postbcd/running-prf.html}, was followed. Briefly, we did a
first pass extraction including the brightest sources only, in order to calculate the
point response function (PRF) for each mosaic. Using this specific PRF we then extracted
sources with a signal-to-noise higher than 5. In the last step, the flux for each source
was estimated by performing PRF fitting. In order to check this procedure, we downloaded
the data from the Great Observatories Origins Deep Survey (GOODS; \citealp{dickinson03})
and compared the results obtained by following this procedure with those reported by the
legacy survey team\footnote{The GOODS Spitzer data can be found at
http://data.spitzer.caltech.edu/popular/goods/}. The flux densities presented in the GOODS
catalog were converted into fluxes assuming a MIPS-24 bandwidth of 4.7~$\mu$m, as reported
by the MIPS data handbook. We found very good agreement between the two independent
studies. On average, the fluxes reported by the GOODS team were higher by $\sim 10\%$ with
a spread of less than $\sim$10\%, even after incorporating an aperture correction. Given
that this small offset will not affect our results, we do not apply any correction to our
fluxes.

\begin{figure}
\begin{center}
\includegraphics[width=0.5\textwidth]{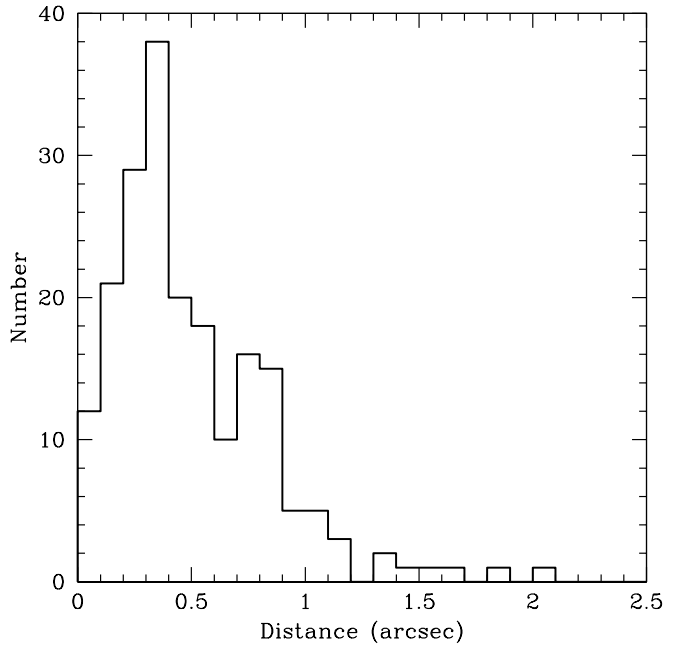}
\end{center}
\caption{Distribution of distances between the optical position from the SDSS and the Spitzer 
MIPS counterpart. Average separation is 0.5$''$.}
\label{dist_hist}
\end{figure}

The SDSS sample was then matched to the sources found in the Spitzer archive, using a
maximum search radius of 10$''$. This maximum nominal separation was chosen both to
minimize the number of false counterparts and to take into account the positional
uncertainties of both Spitzer and SDSS. Altogether, we found in these Spitzer images
counterparts for 205 sources at a signal to noise higher than 5.  In the remaining case,
SDSS J024256.93-001558.0, a very faint counterpart is visible on the Spitzer image, but
not significantly detected so we calculate a 3-$\sigma$ upper limit from the background
standard deviation measured around the SDSS position. The distribution of distances
between the SDSS position and the Spitzer counterpart can be seen in
Figure~\ref{dist_hist}.  The average separation between the optical position and the IR
counterpart was $\sim$0.5$''$.  For each SDSS source, we calculated the probability of a
false match with a 24~$\mu$m counterpart, assuming a Poissonian distribution of sources
and the source density of the corresponding AOR.  We found that the maximum probability of
a false match for a given source was $\sim1.3\%$, with an average probability of 0.02\%.

The observed properties of these 206 sources can be found in Table~\ref{prop_sdss}.  These
objects were on average our most luminous, with bolometric luminosities ranging from $2
\times 10^{45}$~erg~s$^{-1}$ to $3 \times 10^{47}$~erg~s$^{-1}$.  Optical magnitudes in
the AB system were obtained directly from the reported SDSS values. In all cases, the PSF
fitting magnitude was used, as suggested by the SDSS team for sources that are unresolved
at ground-based resolution.  Although the errors in the SDSS photometry are negligible,
since these are all relatively bright optical sources, errors in the MIPS-24~$\mu$m
photometry are estimated to be $\sim$10--$20\%$, including both measurement and systematic
($\sim 5\%$ according to the MIPS data handbook) uncertainties.

\subsection{GOODS Sample Data Collection}

In order to study lower luminosity sources, we include sources detected in both the north
and south GOODS fields.  A summary of the properties of these sources was presented by
\citet{treister06a}.  Our selection criteria were the same as for the SDSS sample:
unobscured sources and 0.8$<$$z$$<$1.2. In this case, a source was classified as
unobscured if broad lines were present in the optical spectrum and the equivalent hydrogen
column density of obscuration $N_H$ in the X-ray spectrum was smaller than
$10^{22}$~cm$^{-2}$.  With these criteria, we found a total of 10 sources, 4 from the
GOODS-N and 6 from the GOODS-S fields. These sources are in general intrinsically fainter
than the SDSS sources, with bolometric luminosities ranging from
$1\times$10$^{44}$~erg~s$^{-1}$ to $1 \times 10^{46}$~erg~s$^{-1}$.

Optical magnitudes for these sources obtained from the Hubble Space telescope with the
Advanced Camera for Surveys (ACS) imager were presented by the GOODS team
\citep{giavalisco04}\footnote{Available at
http://archive.stsci.edu/pub/hlsp/goods/catalog\_r1/}. In order to avoid problems related
to the different spatial resolution of the GOODS and the SDSS images, we used aperture
magnitudes with a 2$''$ diameter to calculate the optical fluxes of the GOODS sources.

The Spitzer properties and redshifts of these sources were obtained from the compilation
of \citet{treister06a}, and are summarized in Table~\ref{prop_goods}. The ID numbers in
table~\ref{prop_goods} correspond to the X-ray identifications from the
\citet{alexander03} catalog. Flux densities in the 24~$\mu$m band were converted into
fluxes by multiplying them by the bandwidth of 4.7~$\mu$m.

\subsection{COSMOS Sample Data Collection}

The COSMOS survey, covering a total area of 2~deg$^2$ at higher flux limits than GOODS,
provides sources that fill the gap in luminosity between the SDSS and GOODS sources.
Their span of luminosities is from $2 \times 10^{44}$~erg~s$^{-1}$ to $1 \times
10^{46}$~erg~s$^{-1}$.  We selected sources from this survey based on the optical
spectroscopy observations of the XMM X-ray sources using the Magellan/IMACS multi-object
spectrograph presented by \citet{trump06}. Using our selection criteria (0.8$<$$z$$<$1.2
and broad optical emission lines) we obtained a sample of 19 sources.

The COSMOS field was completely observed by Spitzer using both IRAC and MIPS as part of
the S-COSMOS legacy program \citep{sanders07}. Using the public catalog of Spitzer sources
generated by the COSMOS team\footnote{Catalog available at
http://irsa.ipac.caltech.edu/Missions/cosmos.html}, we detected 14 of the 19 X-ray
selected sources in the MIPS 24-$\mu$m band. The observed properties of these sources are
presented in Table~\ref{prop_cosmos}. As for the GOODS sources, flux densities in the
24~$\mu$m band were converted into fluxes by multiplying them by the bandwidth.

In summary, our sample consists of 230 sources spanning a range in bolometric luminosity
from $L_{\rm bol}$=10$^{44}$ to 10$^{47.5}$~erg~s$^{-1}$. All the sources in our sample
have measured spectroscopic redshifts and present broad emission lines. In the IR, all
these sources are bright, with luminosities from $L_{\rm IR}$=10$^{43.5}$ to
10$^{46}$~ergs~s$^{-1}$.

\subsection{GALEX Data}

As a large fraction of the bolometric luminosity in unobscured AGN, $\sim$15-20\%, can be
found in the UV (e.g., \citealp{elvis94,richards06,trammell07}), it is worthwhile to look
for UV counterparts of our sources in data taken by the Galaxy Evolution Explorer (GALEX;
\citealp{martin05}).  Specifically, we made use of data release~3\footnote{Available at
http://galex.stsci.edu/GR2/}. In the case of the SDSS sources, we looked for UV
counterparts in a 10$''$ radius around the optical position.  The median separation over
the whole sample was much less than the maximum permitted, $\simeq$0.5$''$, and the actual
maximum separation was only 3.5$''$. 170 of the 206 sources in the SDSS sample were
detected in at least one of the GALEX bands, for a detection rate of 83\%. Of these 170
sources, 61 were detected only in the reddest NUV band and 4 detected only in the FUV
band. The GALEX properties of the SDSS sources are presented in Table~\ref{prop_sdss}.

GALEX coverage of both GOODS fields is significantly deeper.  Whereas the all-sky survey
exposure time was only 0.1~ksec, the North field is being observed for 100~ksec as part of
the deep imaging survey, and the South field is being observed for 200~ksec as part of the
ultra-deep imaging survey.  At the time of data release~2, the total time spent on the
South field was only $\sim$76~ksec, while $\sim$95~ksec had been accumulated in the North
field.  Consequently, only 4 of the 6 sources in the South field were detected by GALEX,
but all the sources in the North field have a UV counterpart.  In Table~\ref{prop_goods},
we present the GALEX properties of the GOODS sources. The entire COSMOS field is one of
the targets of the GALEX Ultra-Deep Imaging Survey. Hence, it is not surprising that all
the sources in our sample with Spitzer-MIPS detections were also detected in the GALEX
observations. The UV properties of the COSMOS sources are presented in
Table~\ref{prop_cosmos}.

\subsection{Bolometric Luminosities}

Because we believe the infrared continuum to be the result of reprocessing shorter
wavelength radiation from the nucleus that strikes the dusty torus, including it in our
estimate of the intrinsic nuclear bolometric luminosity when we have an unobscured view of
the nucleus would be double-counting.  Instead, we do the best we can to sum all flux from
wavelengths shorter than $\sim$1$\mu$m; the exact long wavelength cut-off is not important
because the integral is in general dominated by the UV.  We neglect X-ray contributions,
but these are generally small, $\la 10\%$ (e.g., \citealp{richards06}).

In order to estimate the bolometric luminosities, we combined the photometric information
from GALEX NUV (1750--2800\AA~) to $z$-band ($\sim$1$\mu$m). In the optical range, where
there are only very small gaps in wavelength between the available photometric bands, we
approximate the integrated luminosity by the sum of the observed luminosity in each
band. For UV fluxes, we employ GALEX data wherever possible.  Specifically, we use the
GALEX NUV channel, which for our sample translates to rest-frame wavelengths
875--1400~\AA~.  Whenever the object was either not observed by GALEX, or has only an
upper bound, we estimate the flux in the NUV band by extrapolating from the bluest filter
available ($u$-band for COSMOS and $B$-band for GOODS) using the median quasar SED from
\citet{richards06}.Both because GALEX data are missing from only a small fraction of our
sample and because the intrinsic dispersion in bolometric corrections is $\simeq 50\%$
\citep{richards06}, this extrapolation should not bias significantly our results. To
account for the gap between the $u$ band (rest-frame $\sim 1750$~\AA~) and the GALEX NUV
band, we linearly interpolate in $F_\nu$. At even shorter wavelengths, which are often not
accessible in direct observations, we assume a power-law spectrum ($f_\nu\propto
\nu^\alpha$) with slope $\alpha =-1.76$, as reported by \citet{telfer02} in the
500--1200~\AA~ regime. The amplitude of this power law is fixed by matching to either the
GALEX photometry (when available) or the extrapolated flux from the SDSS composite
spectrum when GALEX data are not available. From the templates of \citet{richards06}, we
estimate that $\sim$20\% of the bolometric luminosity is emitted in this wavelength
range.

The approach described above incorporates all the available information in the optical-UV
range.  In contrast, \citet{maiolino07} computed bolometric luminosities by multiplying
the observed rest-frame 5100~\AA~ luminosity by a fixed bolometric correction.

\section{Trends}

\begin{figure}
\begin{center}
\includegraphics[width=0.5\textwidth]{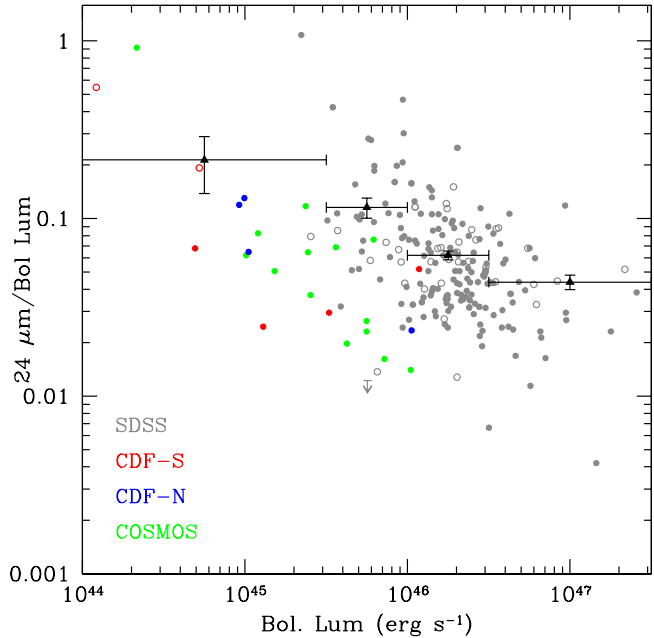}
\end{center}
\caption{Ratio of 24~$\mu$m to bolometric luminosity as a function of
bolometric luminosity for all the sources in the sample. The open circles show
the ratio for sources not detected by GALEX. Triangles with error bars show the
average value in each luminosity bin, with 1-$\sigma$ vertical
error bars.}
\label{24_bol}
\end{figure}

Our central result is shown in Figure~\ref{24_bol}.  There we plot the ratio of MIPS
24~$\mu$m luminosity, $L_{\rm MIPS}$, to our estimate of the bolometric luminosity,
$L_{\rm bol}$, of each individual object.  Two properties of the data are immediately
apparent from this figure: that the mean value of this ratio decreases by order unity over
the factor of 300 range in luminosity these data span; and that at any single luminosity
there is a very large dispersion.

Although $\langle L_{\rm MIPS}/L_{\rm bol}\rangle$ does change over this range of
bolometric luminosities, it is not by a large factor: a factor of 3 is the best
estimate. This figure may be compared with the factor of 2 contrast given by the fitting
formula of \citet{maiolino07}, meant to describe the sample means from $\lambda L_\lambda
(5100$~\AA~$) = 10^{43}$~ergs~s$^{-1}$ to 10$^{47.5}$~ergs~s$^{-1}$, a somewhat greater
dynamic range than spanned by our sample. Nonetheless, our sample size is large enough for
the bin-means to be quite narrowly-defined, and the trend is clearly significant.  Several
statistical tests support this conclusion.  To evaluate the significance in difference
between the mean values in the two bins with the largest number of sample points (the
second and third highest in luminosity), we apply Student's t-test; it shows that they are
different at the $0.1\%$ significance level, while the means of the highest and the third
highest luminosity bins are different at the 10$^{-5}$ level.  Similarly, the Spearman
rank-correlation test finds that the correlation between $\log(L_{\rm bol})$ and
$\log(L_{\rm MIPS}/L_{\rm bol})$ is significant at the 10$^{-7}$ level.  A conventional
Pearson correlation test fails on these data because there are so many more SDSS quasars
than COSMOS and GOODS sources, and the contrast in the mean infrared/bolometric luminosity
across the narrower range of luminosity where most of the SDSS quasars are found is
relatively small.

At the same time, the dispersion is very large.  Considered in logarithmic terms, the rms
fluctuation within the bins is 0.25--0.35, equivalent to a multiplicative factor of
1.8--2.2, comparable to the contrast in the mean across our entire luminosity range.

\begin{figure}
\begin{center}
\includegraphics[width=0.23\textwidth,height=5cm]{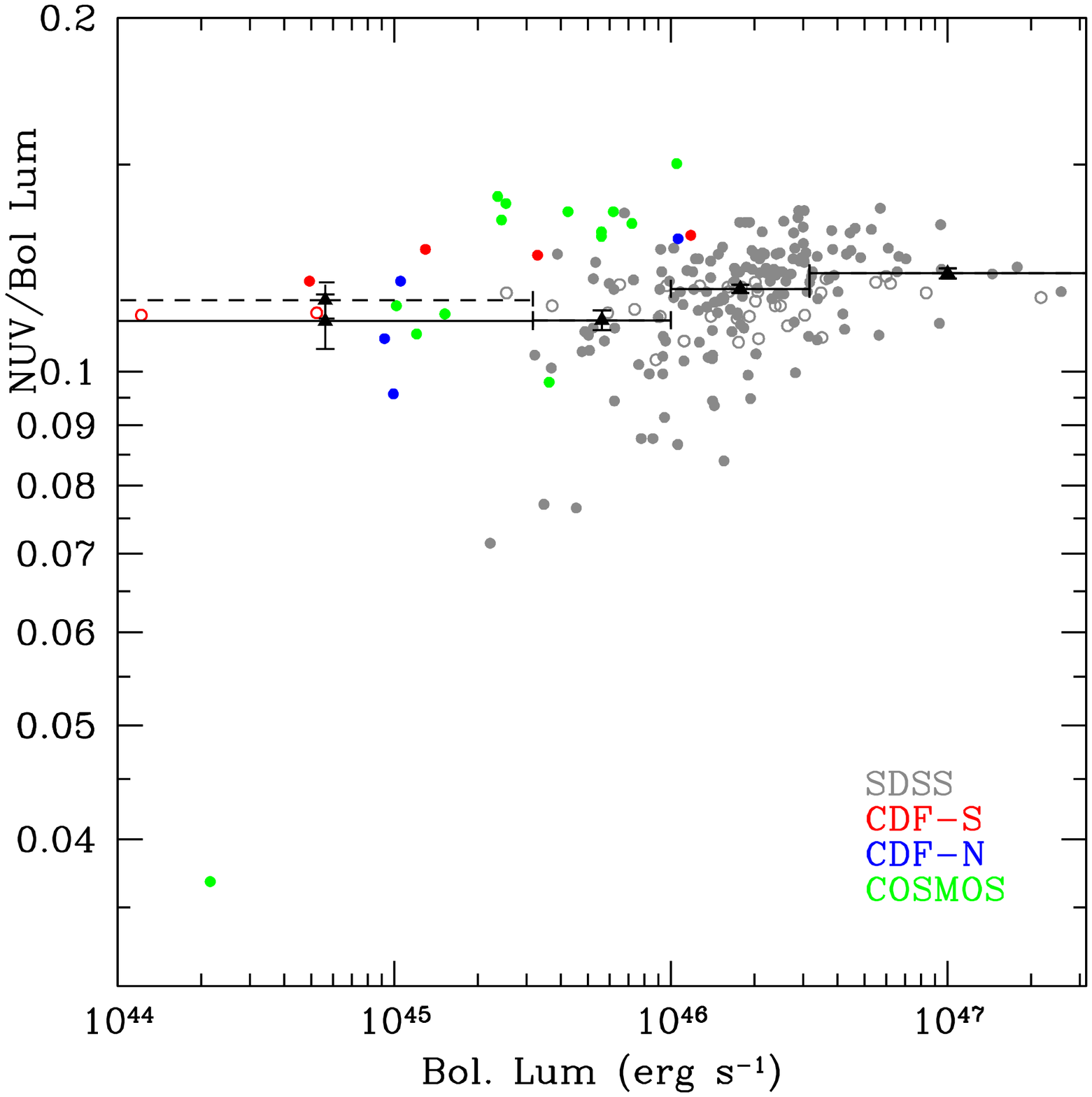}
\includegraphics[width=0.23\textwidth,height=5cm]{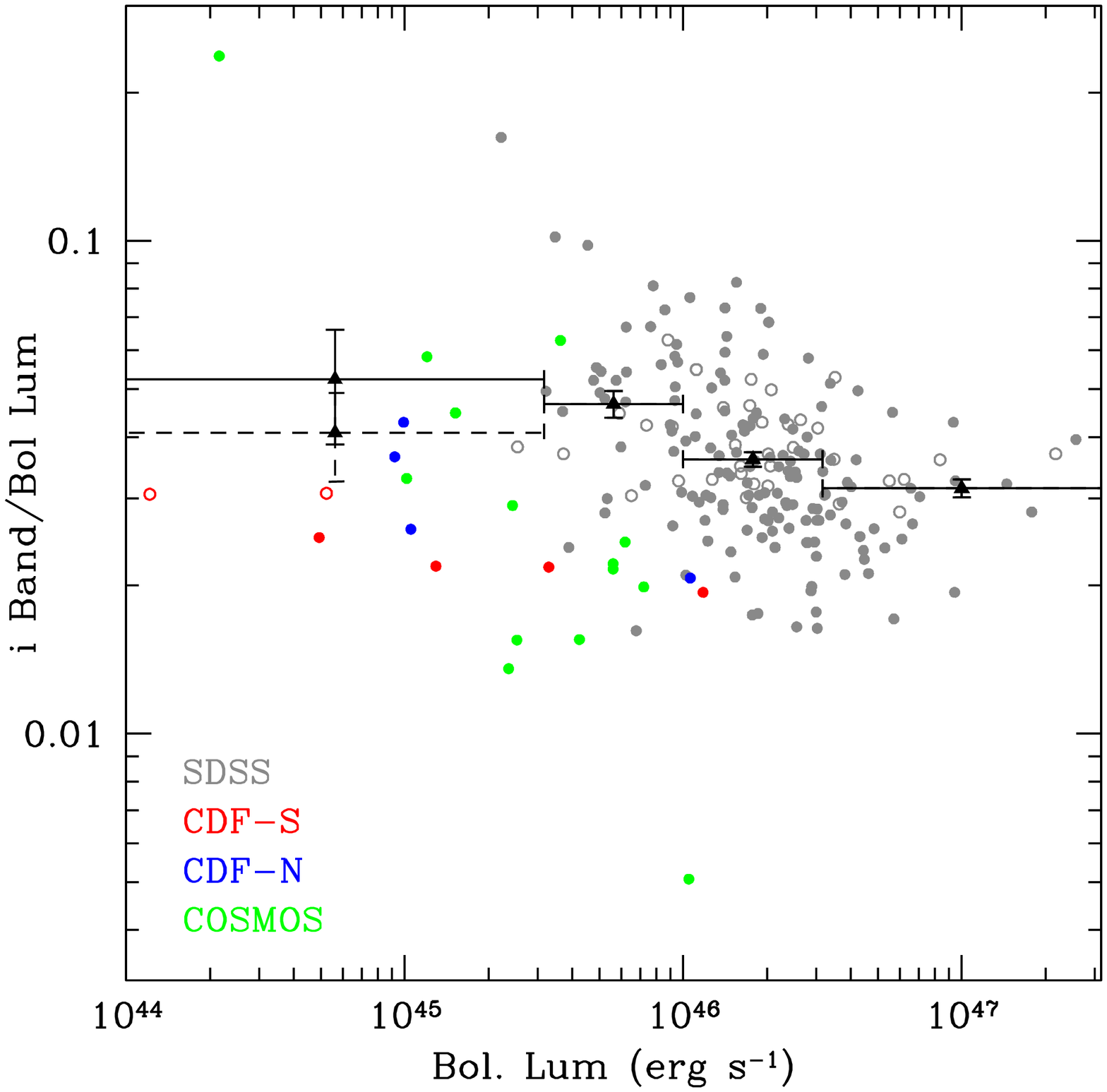}
\end{center}
\caption{Same as Figure~\ref{24_bol} but showing the GALEX NUV to bolometric 
luminosity (left panel) and $i$-band to bolometric luminosity (right panel) ratios as a
function of bolometric luminosity. Dashed error bars show the mean values in the lowest
luminosity bin if source COSMOS J100129.83+023239.0 is excluded.  }
\label{nuv_bol}
\end{figure}

That this trend (and large dispersion) is not related to effects in the intrinsic spectra
of AGN can be seen by considering two other flux ratios as functions of $L_{\rm bol}$
(Fig.~\ref{nuv_bol}). In the mean, there is no significant change in the ratio of either
the $i$-band luminosity or the NUV-band luminosity to bolometric. The small change
observed in the $i$-band to bolometric luminosity ratio is heavily influenced by the
single source COSMOS J100129.83+023239.0, which has a ratio $\sim$8$\times$ higher than
the rest of the sources in that bin. If this source is removed, the average in the lower
luminosity bin decreases from 0.052 to 0.041 and the trend disappears, as can be seen in
Fig.~\ref{nuv_bol}. Performing a Spearman test, we found that the $L_i/L_{\rm bol}$
correlation with $L_{\rm bol}$ is $\sim$3 orders of magnitude less significant than the
$L_{\rm MIPS}/L_{\rm bol}$ correlation and has roughly the same significance as the
$L_{\rm NUV}/L_{\rm bol}$ correlation.

Moreover, there is far less intrinsic scatter at fixed $L_{\rm bol}$ in either of these
ratios than in the infrared to bolometric ratio. The standard deviation of $L_{\rm
IR}/L_{\rm bol}$ in each luminosity bin is a factor of $\sim$2.3, while for $L_i/L_{\rm
bol}$, it is $\sim$1.5 and for $L_{\rm NUV}/L_{\rm bol}$ it is 1.12. The magnitude of the
error induced by using a fixed bolometric correction to estimate bolometric luminosities
from a given optical band is shown by the dispersion of the points plotted in both panels
of Figure~\ref{nuv_bol}.  Use of the i-band and a fixed bolometric correction, for
example, entails an uncertainty in the bolometric luminosity of $\sim 50\%$.

\section{Significance}

\subsection{Evolution?}

Just as it is possible that the physics controlling the geometry of obscuration is
correlated in some way with the bolometric luminosity of the AGN, so too, it may change as
a function of the age of the Universe.  There is evidence, for example, that at fixed
luminosity, $f_{\rm obsc}$ increases with increasing redshift \citep{treister06b}; this
suggestion, however, remains controversial (e.g., \citealp{gilli07}). Factoring out any
possible redshift-dependence was a strong motivating factor for our choice of a sample in
which all the objects have approximately the same $z$.

In fact, the precise character of the correlation between $f_{\rm obsc}$ and $L_{\rm bol}$
also enters into the debate over the possible dependence of $f_{\rm obsc}$ on $z$.
Currently-available samples that probe to high redshift are in general strongly incomplete
at low luminosity because it is so difficult to obtain spectra for objects that faint.  If
one's estimate of the obscured fraction at high redshift is based solely on the detectable
sources (which are, of course, high luminosity objects), it will be biased to a low value
if $f_{\rm obsc}$ systematically declines with increasing luminosity at all redshifts.

\subsection{Relation between $L_{\rm MIPS}/L_{\rm bol}$ and the obscured
solid angle}

As we explained earlier, the ratio of infrared reprocessed light to bolometric luminosity
should increase as the fraction of solid angle occupied by the dusty obscuring gas $f_{\rm
obsc}$ increases.  If the reprocessed IR were radiated isotropically, the ratio
$L_{\rm MIPS}/L_{\rm bol}$ would be proportional to $f_{\rm obsc}$.  It is much more
likely, however, that the reradiation is anisotropic.   Consequently, the relationship
between these two quantities is not necessarily precisely linear, nor
does it directly yield $f_{\rm obsc}$ without the introduction of any additional
parameters (cf. \citealp{maiolino07}).   At the very least (i.e., in the event of
isotropic radiation), there is a proportionality constant to be determined.

At present, there are two lines of evidence regarding the directionality of the reradiated
IR. On the observational side, there are suggestions that it may not be far from isotropic
\citep{lutz04,horst07}.  In both of these studies, it is claimed on the basis of small
samples of AGN that the mean ratio of mid-IR flux to intrinsic hard X-ray flux is the same
for both obscured and unobscured objects.  However, in both of these studies, the
intrinsic hard X-ray luminosities of the obscured objects are in general substantially
smaller (by at least an order of magnitude) than those of the unobscured objects.
Given the correlation we have found between $L_{\rm MIPS}/L_{\rm bol}$ and $L_{\rm bol}$,
one might then have expected that in their samples the mid-IR to X-ray flux ratio would be
even {\it higher} for the obscured than for the unobscured if the IR radiation were
isotropic.  We therefore do not regard these studies as making a strong case for isotropic
radiation.

On the other hand, there are strong physical and theoretical arguments for thinking
that the radiation is anisotropic.  The obscuration is, after all, strongly
aspherical---if it were otherwise, there would be very few unobscured objects to
see.  If the torus is optically thick in the radial direction in the mid-IR and that optical
depth is larger than the vertical optical depth, most of the mid-IR flux should escape
roughly parallel to the torus axis, with rather less escaping in the equatorial direction.
Every detailed radiation transfer model published shows this tendency, although the
precise contrast between the $12\mu$m flux toward the pole and toward the equator varies
depending on the details of the adopted density distribution
\citep{pier93,granato94,efstathiou95,nenkova02,vanbemmel03,dullemond05}.  The most recent
paper (which presents both smooth and clumpy models) finds ratios between the total
infrared flux in the equatorial plane and at $20^{\circ}$ from the axis to vary between
0.06 and 0.3, depending on the specific model.  Making the approximation that very little
flux is directed toward observers in the obscured solid angle, we expect that the
MIPS-band luminosity should scale relative to the bolometric luminosity approximately as
\begin{equation}
\frac{dL_{\rm MIPS}}{d\Omega} \simeq f_{12}(\theta)\frac{L_{\rm bol}}{4\pi}
                               \frac{f_{\rm obsc}}{1 - f_{\rm obsc}}.
\end{equation}
Here $f_{12}(\theta)$ is the fraction of the total dust-reprocessed luminosity falling
within the MIPS band ($\simeq 11$--$13\mu$m in the rest-frame at $z \simeq 1$).  Even
within the range of viewing angles $\theta$ that provide an unobscured view, we may expect
order unity variations in this quantity.

To confirm this scaling and estimate $f_{12}(\theta)$, we studied the results of a number
of model calculations.  These models were performed with the code described in
\cite{dullemond05}.  For all of them, the obscuration was assumed to be confined to within
a constant opening-angle spherical wedge whose outer radius was 30 times greater than its
inner radius.  We considered two density profiles: a single constant value and a radial
decay $\propto$$r^{-1}$. For both density models, when the torus is optically thick in the
mid-infrared, the approximate model given by equation~1 is confirmed provided the opening
angle is at least $\simeq$0.7~radians. Crudely speaking, the solid angle-weighted mean
value of $f_{12}$ (for viewing angles permitting an unobscured view of the nucleus) does
not vary a great deal with torus opening angle: it varies from $\simeq$0.06 to
$\simeq$0.08 depending on the model details for the same range of opening angles for which
equation~1 holds.  When the opening angle becomes smaller than this, $f_{12}$ can begin to
diminish, with the amount sensitive to whether the torus is truly a spherical wedge or has
a flatter outer envelope.  At fixed opening angle, $f_{12}$ does vary somewhat with
viewing angle, typically being perhaps $50\%$ greater than the mean for nearly polar
viewing angle and dropping sharply when the viewing angle becomes close to the torus
opening angle.  The latter effect is also sensitive to the specific geometry (spherical
wedge) adopted in these models, which also depends slightly on the wavelength of the
emission.

The ratio of IR flux to bolometric seen by an observer in the unobscured direction is then
\begin{equation}
\frac{L_{\rm MIPS}}{L_{\rm bol}} \simeq f_{12}(\theta)\frac{f_{\rm obsc}}
                    {1 - f_{\rm obsc}}.
\end{equation}
To this level of approximation, the obscured fraction associated with a given value of
$L_{\rm MIPS}/L_{\rm bol}$ is

\begin{equation}\label{eq:fobsc}
f_{\rm obsc} \simeq \frac{1}{1 + q},
\end{equation}
where 
\begin{equation}
q = \frac{f_{12}(\theta)}{L_{\rm MIPS}/L_{\rm bol}}.
\end{equation}

\begin{figure}
\begin{center}
\includegraphics[width=0.5\textwidth]{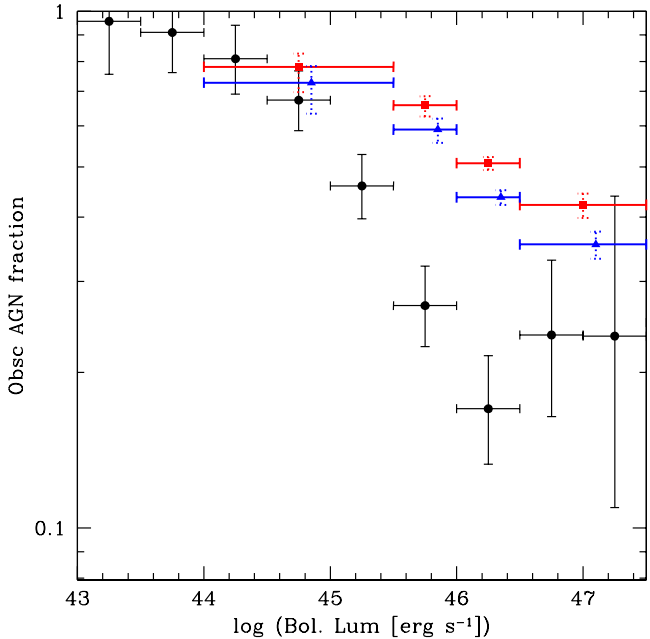}
\end{center}
\caption{Two different inferences of $f_{\rm obsc}$: black curve with
error bars from X-ray data, as described in the text; triangles with error bars from the
method of this paper assuming values of $f_{12}$=0.06 (squares; red in electronic version) and 0.08 (triangles; blue in electronic version; displaced by
$\log$$L_{\rm bol}$=0.1 for clarity). The horizontal error bars simply
describe the width of the luminosity bin; the vertical red and blue error bars are dashed
to indicate that they are model-dependent in the sense discussed in the text.  Their
extent was determined by translating the error bars of Fig.~\ref{24_bol} using
Eqn.~\ref{eq:fobsc}.}
\label{obs_frac}
\end{figure}

Using the solid angle-weighted means of $f_{12}$ taken from our models, we
can transform the bin-means of $L_{\rm MIPS}/L_{\rm bol}$ shown in Figure~\ref{24_bol}
to estimates of $f_{\rm obsc}$.  The results are shown in Figure~\ref{obs_frac}.
It appears that the mean value of $f_{\rm obsc}$ falls from $\simeq 0.9$ at the
lowest bolometric luminosities in our sample to $\simeq 0.3$--0.4 at the highest
luminosities.  {\it We strongly caution, however, that there may be a systematic
error of order unity in $\langle f_{12}\rangle$ due to the specific assumptions made in
those models regarding the density distribution within the envelope that determines
$f_{\rm obsc}$.  A variety of parameters (e.g., the ratio of outer radius to inner,
the degree of inhomogeneity and clumping, etc.) may quantitatively alter
$\langle f_{12}\rangle$.}

We compare this dependence of the mean $f_{\rm obsc}$ as a function of $L_{\rm bol}$ with
the relation that has been found in X-ray studies in Figure~\ref{obs_frac}.  For this
figure we use the X-ray sample of \citet{treister06b}, with the X-ray luminosity
transformed to $L_{\rm bol}$ using the luminosity-dependent bolometric corrections of
\citet{marconi04}. The level of agreement between the two independent methods of inferring
$f_{\rm obsc}$ is surprisingly good, considering the systematic errors in both.  Most
notably, X-ray surveys completely miss objects with column densities $\gg
10^{24}$~cm$^{-2}$, and there are indications that these objects may be numerous
\citep{martinez06,polletta06}.  Indeed, if one wished to extend credence to the somewhat
larger $f_{\rm obsc}$ found by the infrared method at intermediate luminosities, it might
be explained by a somewhat larger proportion of Compton-thick objects within the
population. It is also interesting that, as pointed out by \citet{maiolino07}, the values
of $f_{\rm obsc}$ derived from X-ray surveys include the effects of obscuration by both
dust and gas, while the IR-derived values include only obscuration by dust.

To close this subsection, we remark on another distinction that may be important in
evaluating the meaning of the $f_{\rm obsc}$ estimated from $L_{\rm MIPS}/L_{\rm bol}$.
It is possible for obscuring matter of smaller optical depth to intercept some of the
nuclear light farther out in the host galaxy and reradiate it in the mid-infrared even
though this matter is wholly independent of the ``obscuring torus" proper (e.g.,
\citealp{keel80,mcleod95,rigby06}).  Molecular clouds and other gas concentrations in the
host have at most modest column densities when measured in terms of X-ray absorption
(generally $< 10^{22}$~cm$^{-2}$), but such column densities can still absorb large
portions of an AGN's optical/UV continuum.  Consequently, they qualify as contributing to
$f_{\rm obsc}$ both in the sense that they can block our view of the optical/UV light from
the nucleus and, depending on their temperature, in terms of their contribution to the
mid-infrared luminosity of the system. 

\subsection{The dispersion at fixed luminosity}

One of the virtues of using SDSS quasars is that the sample size, particularly for $5
\times 10^{45}$~erg~s$^{-1} < L_{\rm bol} < 1 \times 10^{47}$~erg~s$^{-1}$, is large
enough to be truly statistical.  As a result, we can obtain a reasonably reliable estimate
of the sample dispersion as well as its mean.  As we have previously remarked, the
dispersion is quite large.

We see no measurement error that could contribute in a significant way to the dispersion
in $L_{\rm MIPS}/L_{\rm bol}$ at fixed $L_{\rm bol}$. The errors in the photometry we use
to compute the bolometric luminosity are small, $\sim$2\%; the uncertainty in the MIPS
fluxes is $\sim$10-20\%, as detailed in \S 2.1.

Intrinsic variability could also contribute, but in a similarly minor fashion except
perhaps in a very small number of objects.  The time interval between the GALEX
measurements, the SDSS photometry, and the Spitzer measurements was generally a few years.
On that timescale, SDSS data \citep{vandenberk04} show that quasars typically vary by
$0.1$--0.2~mag in the optical, with the larger variability usually occurring at shorter
wavelengths and in lower luminosity objects.  A small fraction vary by as much as $\simeq
0.5$~mag on these timescales, still much smaller than our observed dispersion relative to
the mean.  If the infrared is, as we assume, the result of thermal reprocessing by dust,
it cannot vary substantially on timescales shorter than several to ten years.

According to the results of \citet{schweitzer06}, the contribution of star formation to
the total light in quasars is $\sim 30\%$ at far-IR wavelengths, in agreement with the
results of \citet{shi07}, who found a contribution of $\sim 25\%$ at 70~$\mu$m and
$\sim 10\%$ at 24~$\mu$m. Hence, extrapolating linearly to the MIPS band, we expect a
contribution of star formation to the total IR light of $\sim5\%$ at rest-frame
12~$\mu$m; we therefore do not expect our results to be significantly affected by star
formation processes in the host galaxy. In addition, because $f_{obsc}$ is larger for
lower luminosity sources than for higher luminosity sources and the nuclear infrared
luminosity is $f_{obsc}L_{\rm bol}$, we do not expect the contribution from star
formation to increase significantly even at lower luminosities.

There must, therefore, be a genuine large dispersion in $L_{\rm MIPS}/L_{\rm bol}$.  One
possible source of this dispersion is the variation of $f_{12}$ with viewing angle.  Again
relying on the models described above, a total range of a factor of 2--3 may be expected
from this source alone.  Compared with our rms logarithmic dispersion of a factor of
$\sim$2, or the factor of 100 total range seen in the most-populated bins, the expected
variations in viewing angle can account for a significant part of this dispersion, but
likely not all of it.

This last statement must, of course, be qualified by the condition that our models
accurately estimate the range in $f_{12}(\theta)$.  It is entirely possible that changes
in the detailed density distribution, even while keeping $f_{\rm obsc}$ fixed, might
introduce further variations in $f_{12}(\theta)$.  Although the expected large optical
depth in the $12\mu$m band is likely to limit the sensitivity of $f_{12}(\theta)$ to
changes in the radial density profile or density inhomogeneities, there could certainly be
small, but detectable, quantitative effects.
    
Finally, there may, of course, be additional dispersion due to variations in $f_{\rm
obsc}$ at fixed luminosity. It is entirely possible for parameters other than $L_{\rm
bol}$ to influence the obscuration's solid angle.  Among the obvious candidates are the
mass of the central mass black hole and the stellar mass contained within the region of
the obscuring matter.  The varying impact of magnetic forces, clump collisions, or other
mechanisms could also conceivably play a part.

\subsection{Implications for surveys}

Particularly since the launch of the {\it Spitzer Space Telescope}, it has become
feasible---and worthwhile---to construct samples of AGN based on mid-infrared selection
\citep{lacy04,stern05,martinez06}.  Our results here demonstrate that the translation
between mid-infrared flux and bolometric flux has an intrinsic scatter of a factor of
$\simeq 2$ up or down, even for unobscured AGN.  Because, as we have discussed, $L_{\rm
IR}/L_{\rm bol}$ is generally smaller in obscured AGN than in unobscured, the bolometric
correction based on infrared flux for them should be an even larger factor, with perhaps
greater scatter.

\section{Conclusions}

We presented in this paper an alternative way to estimate the fraction of obscured AGN and
its possible dependence on luminosity, by computing the relative fraction of IR emission
in unobscured AGN spanning a large of luminosities. Using a sample of 206
optically-selected high luminosity AGN from the SDSS and 24 X-ray selected lower
luminosity AGN from the GOODS and COSMOS surveys with 0.8$\leq$$z$$\leq$1.2, nearly all
having Spitzer detections at 24~$\mu$m, we found a decrease of a factor of $\sim$3 in the
relative IR emission at rest-frame 12~$\mu$m from $L_{\rm bol}$=10$^{44}$~ergs~s$^{-1}$ to
10$^{47.5}$~ergs~s$^{-1}$. In addition, we found a significant scatter of a factor of 2--3
at a given bolometric luminosity. Some of this scatter can likely be explained by a
dependence of the 11--13~$\mu$m fraction $f_{12}$ on viewing angle; some may also be due
to an intrinsic scatter in $f_{\rm obsc}$ at fixed bolometric luminosity due to a
dependence on other parameters such as black hole mass, magnetic forces, etc.

Using IR re-emission models in order to convert $L_{\rm MIPS}/L_{\rm bol}$ into a fraction
of obscured AGN, we found that $f_{\rm obsc}$ changes from $\sim 90\%$ at $L_{\rm bol} =
10^{44}$~ergs~s$^{-1}$ to $\sim 30$--$40\%$ at $L_{\rm bol} \simeq
10^{47}$~ergs~s$^{-1}$. The derived dependence of the obscured fraction on luminosity is
in good agreement with the direct, but possibly biased (by the omission of Compton-thick
objects) observation of this fraction in X-ray surveys.  We caution, however, that there
may be systematic error in the precise values of $f_{\rm obsc}$ inferred from $L_{\rm
MIPS}/L_{\rm bol}$ because our estimate of $\langle f_{12}\rangle$ is likely to depend
somewhat on the details of the density distribution within the torus.  Nonetheless, it is
interesting that in general the values of $f_{\rm obsc}$ we infer are larger than what is
seen in X-ray surveys.  This contrast hints that there may be a significant population of
heavily obscured, even Compton thick, AGN that are missed in X-ray observations but
included in our samples.

\acknowledgements

We thank ESO-Chile for JHK's support through its senior visitor program, and the
hospitality given him during his visit to Santiago. We thank the anonymous referee for
very useful comments and suggestions. This work is based in part on archival data obtained
with the Spitzer Space Telescope, which is operated by the Jet Propulsion Laboratory,
California Institute of Technology under a contract with NASA.  It was partially supported
by a Spitzer Archival Research grant, subcontract number 1310126 from Caltech/JPL.
Funding for the SDSS and SDSS-II has been provided by the Alfred P. Sloan Foundation, the
Participating Institutions, the National Science Foundation, the U.S. Department of
Energy, the National Aeronautics and Space Administration, the Japanese Monbukagakusho,
the Max Planck Society, and the Higher Education Funding Council for England. The SDSS Web
Site is http://www.sdss.org/.

\clearpage\thispagestyle{empty}
\begin{deluxetable}{lccccccccccccccc}
\tablecolumns{16}
\tabletypesize{\scriptsize}
\tablewidth{0pc}
\tablecaption{\label{prop_sdss} Observed Properties of Sources in the SDSS Sample}
\tablehead{
\colhead{Name} & \colhead{Redshift} & \multicolumn{5}{c}{Optical Mag. (AB)} & \multicolumn{7}{c}{Flux (10$^{-12}$~erg~cm$^{-2}$~s$^{-1}$)} & \colhead{log(Bol. Lum)}\\
\colhead{} & \colhead{} & \colhead{$u$} & \colhead{$g$} & \colhead{$r$} & \colhead{$i$} & \colhead{$z$} & \colhead{24~$\mu$m} & \multicolumn{3}{c}{NUV} & \multicolumn{3}{c}{FUV} & \colhead{erg~s$^{-1}$}\\
\colhead{} &\colhead{} & \colhead{}    & \colhead{}    & \colhead{}    & \colhead{}    & \colhead{}    &  \colhead{flux} & \colhead{flux} & \colhead{upper} & \colhead{lower} & \colhead{flux} & \colhead{upper} & \colhead{lower} & \colhead{}
}
\startdata
SDSS J110116.41+572850.5 & 0.80 & 17.85 & 17.53 & 17.48 & 17.46 & 17.40 & 0.331 & 1.335 & 1.341 & 1.332 & 0.466 & 0.470 & 0.462 & 46.5\\
SDSS J142810.31+353847.0 & 0.80 & 19.09 & 18.79 & 18.74 & 18.89 & 18.71 & 0.120 & 0.431 & 0.436 & 0.426 & 0.179 & ----- & ----- & 46.0\\
SDSS J104755.02+120850.2 & 0.81 & 19.84 & 19.43 & 19.16 & 19.02 & 18.71 & 0.537 & 0.202 & 0.214 & 0.191 & 0.012 & 0.016 & 0.008 & 45.8\\
SDSS J143345.10+345939.9 & 0.81 & 20.36 & 19.64 & 19.10 & 18.84 & 18.49 & 0.485 & 0.089 & 0.091 & 0.086 & 0.055 & ----- & ----- & 45.5\\
SDSS J160630.60+542007.5 & 0.82 & 19.03 & 18.77 & 18.73 & 18.83 & 18.61 & 0.413 & 0.455 & 0.459 & 0.452 & 0.185 & ----- & ----- & 46.1\\
SDSS J235948.53-103938.5 & 0.83 & 18.07 & 17.87 & 17.79 & 17.80 & 17.71 & 0.439 & 1.209 & 1.226 & 1.190 & 0.421 & 0.436 & 0.407 & 46.5\\
\enddata
\tablecomments{This table is published in its entirety in the electronic edition of the Astrophysical Journal. A portion is shown here for guidance regarding its form and content.}
\end{deluxetable}

\begin{deluxetable}{lccccccccccccccc}
\tablecolumns{16}
\tabletypesize{\scriptsize}
\tablewidth{0pc}
\tablecaption{\label{prop_goods} Observed Properties of Sources in the GOODS Sample}
\tablehead{
\colhead{Field} & \colhead{ID} & \colhead{Redshift} & \multicolumn{4}{c}{Optical Mag. (AB)} & \multicolumn{7}{c}{Flux (10$^{-12}$~erg~cm$^{-2}$~s$^{-1}$)} & \colhead{log(Bol. Lum)}\\
\colhead{} & \colhead{} & \colhead{} & \colhead{$B$} & \colhead{$V$} & \colhead{$i$} & \colhead{$z$} & \colhead{24~$\mu$m} & \multicolumn{3}{c}{NUV} & \multicolumn{3}{c}{FUV} & \colhead{erg~s$^{-1}$}\\
\colhead{} & \colhead{} & \colhead{} &  \colhead{} & \colhead{} & \colhead{} & \colhead{} & \colhead{flux} & \colhead{flux} & \colhead{upper} & \colhead{lower} & \colhead{flux} & \colhead{upper} & \colhead{lower} & \colhead{}}
\startdata
South & 34 & 1.040 & 22.72 & 22.36 & 22.23 & 21.23 & 0.006 & 0.029 & 0.030 & 0.028 & 0.006 & 0.007 & 0.006 & 45.1\\
South & 173 & 1.030 & 20.17 & 20.00 & 19.94 & 19.03 & 0.112 & 0.281 & 0.283 & 0.280 & 0.092 & 0.093 & 0.091 & 46.1\\
South & 214 & 0.840 & 24.09 & 23.47 & 22.56 & 21.34 & 0.010 & 0.018 & 0.020 & 0.016 & 0.009 & 0.011 & 0.008 & 44.7\\
South & 234 & 0.840 & 20.95 & 20.54 & 20.65 & 19.65 & 0.029 & 0.125 & 0.126 & 0.124 & 0.038 & 0.039 & 0.037 & 45.5\\
South & 193 & 0.960 & 23.71 & 23.16 & 22.63 & 21.40 & 0.022 & 0.013 & ----- & ----- & 0.005 & ----- & ----- & 44.7\\
South & 200 & 0.960 & 25.49 & 24.89 & 24.22 & 22.92 & 0.014 & 0.003 & ----- & ----- & 0.001 & ----- & ----- & 44.1\\
North & 116 & 1.022 & 22.62 & 22.08 & 22.00 & 20.88 & 0.021 & 0.018 & 0.019 & 0.018 & 0.008 & ----- & ----- & 45.0\\
North & 340 & 0.903 & 22.91 & 22.20 & 21.89 & 20.75 & 0.017 & 0.032 & 0.033 & 0.031 & 0.009 & ----- & ----- & 45.0\\
North & 344 & 1.018 & 20.13 & 19.89 & 19.95 & 19.15 & 0.047 & 0.259 & 0.260 & 0.258 & 0.050 & ----- & ----- & 46.0\\
North & 451 & 0.837 & 28.47 & 21.33 & 21.21 & 20.09 & 0.039 & 0.029 & 0.030 & 0.028 & 0.019 & ----- & ----- & 45.0\\
\enddata
\end{deluxetable}


\begin{deluxetable}{lccccccccccccccc}
\tablecolumns{16}
\tabletypesize{\scriptsize}
\tablewidth{0pc}
\tablecaption{\label{prop_cosmos} Observed Properties of Sources in the COSMOS Sample}
\tablehead{
\colhead{Name} & \colhead{Redshift} & \multicolumn{5}{c}{Optical Mag. (AB)} & \multicolumn{7}{c}{Flux (10$^{-12}$~erg~cm$^{-2}$~s$^{-1}$)} & \colhead{log(Bol. Lum)}\\
\colhead{} & \colhead{} & \colhead{$u$} & \colhead{$g$} & \colhead{$r$} & \colhead{$i$} & \colhead{$z$} & \colhead{24~$\mu$m} & \multicolumn{3}{c}{NUV} & \multicolumn{3}{c}{FUV} & \colhead{erg~s$^{-1}$}\\
\colhead{} &\colhead{} & \colhead{}    & \colhead{}    & \colhead{}    & \colhead{}    & \colhead{}    &  \colhead{flux} & \colhead{flux} & \colhead{upper} & \colhead{lower} & \colhead{flux} & \colhead{upper} & \colhead{lower} & \colhead{}
}
\startdata
COSMOS J095902.56+022511.8 & 1.105 & 22.39 & 22.69 & 22.42 & 22.23 & 21.06 & 0.010 & 0.018 & 0.021 & 0.015 & 0.007 & ----- & ----- & 45.0\\
COSMOS J095940.06+022306.8 & 1.132 & 20.92 & 20.62 & 20.29 & 20.22 & 20.35 & 0.036 & 0.051 & 0.078 & 0.034 & 0.026 & ----- & ----- & 45.6\\
COSMOS J095946.92+022209.5 & 0.909 & 22.07 & 21.87 & 21.64 & 21.76 & 22.71 & 0.069 & 0.083 & 0.084 & 0.082 & 0.026 & 0.027 & 0.025 & 45.4\\
COSMOS J100033.38+015237.2 & 0.831 & 21.17 & 20.83 & 20.74 & 20.73 & 20.07 & 0.026 & 0.180 & 0.215 & 0.152 & 0.026 & ----- & ----- & 45.6\\
COSMOS J100034.93+020235.2 & 1.177 & 21.14 & 21.40 & 21.05 & 20.98 & 20.51 & 0.020 & 0.096 & 0.097 & 0.095 & 0.030 & 0.031 & 0.029 & 45.7\\
COSMOS J100042.37+014534.1 & 1.161 & 22.54 & 22.88 & 22.27 & 21.86 & 20.84 & 0.020 & 0.215 & 0.275 & 0.167 & 0.006 & ----- & ----- & 46.0\\
COSMOS J100049.97+015231.3 & 1.156 & 21.21 & 21.27 & 21.00 & 20.96 & 20.65 & 0.018 & 0.101 & 0.103 & 0.100 & 0.038 & 0.039 & 0.037 & 45.7\\
COSMOS J100114.86+020208.8 & 0.989 & 21.91 & 21.82 & 21.42 & 21.13 & 21.12 & 0.020 & 0.026 & 0.042 & 0.016 & 0.011 & ----- & ----- & 45.1\\
COSMOS J100118.57+022739.4 & 1.052 & 22.07 & 21.26 & 20.64 & 20.46 & 20.24 & 0.082 & 0.147 & 0.208 & 0.104 & 0.009 & ----- & ----- & 45.8\\
COSMOS J100129.83+023239.0 & 0.825 & 24.80 & 22.72 & 21.66 & 20.99 & 20.88 & 0.062 & 0.003 & ----- & ----- & 0.004 & 0.005 & 0.003 & 44.3\\
COSMOS J100141.33+021031.5 & 0.982 & 22.29 & 21.96 & 21.39 & 21.10 & 21.44 & 0.032 & 0.068 & 0.069 & 0.066 & 0.014 & 0.015 & 0.013 & 45.4\\
COSMOS J100151.11+020032.7 & 0.964 & 20.59 & 20.49 & 20.26 & 20.29 & 20.21 & 0.025 & 0.208 & 0.212 & 0.203 & 0.079 & 0.082 & 0.076 & 45.9\\
COSMOS J100159.86+013135.3 & 0.977 & 22.31 & 21.96 & 21.73 & 21.73 & 21.45 & 0.020 & 0.073 & 0.075 & 0.072 & 0.037 & 0.038 & 0.036 & 45.4\\
COSMOS J100229.33+014528.1 & 0.876 & 21.90 & 21.25 & 20.77 & 20.84 & 20.39 & 0.021 & 0.047 & 0.047 & 0.046 & 0.027 & 0.028 & 0.026 & 45.2\\
\enddata
\end{deluxetable}


\begin{thebibliography}{55}
\expandafter\ifx\csname natexlab\endcsname\relax\def\natexlab#1{#1}\fi

\bibitem[{{Adelman-McCarthy} {et~al.}(2006)}]{adelman06}
{Adelman-McCarthy}, J.~K. {et~al.} 2006, \apjs, 162, 38

\bibitem[{{Alexander} {et~al.}(2003){Alexander}, {Bauer}, {Brandt},
  {Schneider}, {Hornschemeier}, {Vignali}, {Barger}, {Broos}, {Cowie},
  {Garmire}, {Townsley}, {Bautz}, {Chartas}, \& {Sargent}}]{alexander03}
{Alexander}, D.~M., {Bauer}, F.~E., {Brandt}, W.~N., {Schneider}, D.~P.,
  {Hornschemeier}, A.~E., {Vignali}, C., {Barger}, A.~J., {Broos}, P.~S.,
  {Cowie}, L.~L., {Garmire}, G.~P., {Townsley}, L.~K., {Bautz}, M.~W.,
  {Chartas}, G., \& {Sargent}, W.~L.~W. 2003, \aj, 126, 539

\bibitem[{{Alonso-Herrero} {et~al.}(2006){Alonso-Herrero},
  {P{\'e}rez-Gonz{\'a}lez}, {Alexander}, {Rieke}, {Rigopoulou}, {Le Floc'h},
  {Barmby}, {Papovich}, {Rigby}, {Bauer}, {Brandt}, {Egami}, {Willner}, {Dole},
  \& {Huang}}]{alonso-herrero06}
{Alonso-Herrero}, A., {P{\'e}rez-Gonz{\'a}lez}, P.~G., {Alexander}, D.~M.,
  {Rieke}, G.~H., {Rigopoulou}, D., {Le Floc'h}, E., {Barmby}, P., {Papovich},
  C., {Rigby}, J.~R., {Bauer}, F.~E., {Brandt}, W.~N., {Egami}, E., {Willner},
  S.~P., {Dole}, H., \& {Huang}, J.-S. 2006, \apj, 640, 167

\bibitem[{{Antonucci}(1993)}]{antonucci93}
{Antonucci}, R. 1993, \araa, 31, 473

\bibitem[{{Antonucci} \& {Miller}(1985)}]{antonucci85}
{Antonucci}, R.~R.~J. \& {Miller}, J.~S. 1985, \apj, 297, 621

\bibitem[{{Barger} {et~al.}(2005){Barger}, {Cowie}, {Mushotzky}, {Yang},
  {Wang}, {Steffen}, \& {Capak}}]{barger05}
{Barger}, A.~J., {Cowie}, L.~L., {Mushotzky}, R.~F., {Yang}, Y., {Wang}, W.-H.,
  {Steffen}, A.~T., \& {Capak}, P. 2005, \aj, 129, 578

\bibitem[{{di Serego Alighieri} {et~al.}(1994){di Serego Alighieri}, {Cimatti},
  \& {Fosbury}}]{diserego94}
{di Serego Alighieri}, S., {Cimatti}, A., \& {Fosbury}, R.~A.~E. 1994, \apj,
  431, 123

\bibitem[{{Dickinson} {et~al.}(2003){Dickinson}, {Giavalisco}, \& {The Goods
  Team}}]{dickinson03}
{Dickinson}, M., {Giavalisco}, M., \& {The Goods Team}. 2003, in The Mass of
  Galaxies at Low and High Redshift, ed. R.~{Bender} \& A.~{Renzini}, 324--+

\bibitem[{{Dullemond} \& {van Bemmel}(2005)}]{dullemond05}
{Dullemond}, C.~P. \& {van Bemmel}, I.~M. 2005, \aap, 436, 47

\bibitem[{{Efstathiou} \& {Rowan-Robinson}(1995)}]{efstathiou95}
{Efstathiou}, A. \& {Rowan-Robinson}, M. 1995, \mnras, 273, 649

\bibitem[{{Elvis} {et~al.}(1994){Elvis}, {Wilkes}, {McDowell}, {Green},
  {Bechtold}, {Willner}, {Oey}, {Polomski}, \& {Cutri}}]{elvis94}
{Elvis}, M., {Wilkes}, B.~J., {McDowell}, J.~C., {Green}, R.~F., {Bechtold},
  J., {Willner}, S.~P., {Oey}, M.~S., {Polomski}, E., \& {Cutri}, R. 1994,
  \apjs, 95, 1

\bibitem[{{Ferruit} {et~al.}(2000){Ferruit}, {Wilson}, \&
  {Mulchaey}}]{ferruit00}
{Ferruit}, P., {Wilson}, A.~S., \& {Mulchaey}, J. 2000, \apjs, 128, 139

\bibitem[{{Giavalisco} {et~al.}(2004)}]{giavalisco04}
{Giavalisco}, M. {et~al.} 2004, \apjl, 600, L93

\bibitem[{{Gilli} {et~al.}(2007){Gilli}, {Comastri}, \& {Hasinger}}]{gilli07}
{Gilli}, R., {Comastri}, A., \& {Hasinger}, G. 2007, \aap, 463, 79

\bibitem[{{Granato} \& {Danese}(1994)}]{granato94}
{Granato}, G.~L. \& {Danese}, L. 1994, \mnras, 268, 235

\bibitem[{{Hao} {et~al.}(2005){Hao}, {Strauss}, {Fan}, {Tremonti}, {Schlegel},
  {Heckman}, {Kauffmann}, {Blanton}, {Gunn}, {Hall}, {Ivezi{\'c}}, {Knapp},
  {Krolik}, {Lupton}, {Richards}, {Schneider}, {Strateva}, {Zakamska},
  {Brinkmann}, \& {Szokoly}}]{hao05}
{Hao}, L., {Strauss}, M.~A., {Fan}, X., {Tremonti}, C.~A., {Schlegel}, D.~J.,
  {Heckman}, T.~M., {Kauffmann}, G., {Blanton}, M.~R., {Gunn}, J.~E., {Hall},
  P.~B., {Ivezi{\'c}}, {\v Z}., {Knapp}, G.~R., {Krolik}, J.~H., {Lupton},
  R.~H., {Richards}, G.~T., {Schneider}, D.~P., {Strateva}, I.~V., {Zakamska},
  N.~L., {Brinkmann}, J., \& {Szokoly}, G.~P. 2005, \aj, 129, 1795

\bibitem[Horst {et~al.}(2007)]{horst07}
  Horst, H., Gandhi, P., Smette, A. \& Duschl, W.J. 2007, A\&A in press, arXiv:0711.3734

\bibitem[{{Jaffe} {et~al.}(2004){Jaffe}, {Meisenheimer}, {R{\"o}ttgering},
  {Leinert}, {Richichi}, {Chesneau}, {Fraix-Burnet}, {Glazenborg-Kluttig},
  {Granato}, {Graser}, {Heijligers}, {K{\"o}hler}, {Malbet}, {Miley},
  {Paresce}, {Pel}, {Perrin}, {Przygodda}, {Schoeller}, {Sol}, {Waters},
  {Weigelt}, {Woillez}, \& {de Zeeuw}}]{jaffe04}
{Jaffe}, W., {Meisenheimer}, K., {R{\"o}ttgering}, H.~J.~A., {Leinert}, C.,
  {Richichi}, A., {Chesneau}, O., {Fraix-Burnet}, D., {Glazenborg-Kluttig}, A.,
  {Granato}, G.-L., {Graser}, U., {Heijligers}, B., {K{\"o}hler}, R., {Malbet},
  F., {Miley}, G.~K., {Paresce}, F., {Pel}, J.-W., {Perrin}, G., {Przygodda},
  F., {Schoeller}, M., {Sol}, H., {Waters}, L.~B.~F.~M., {Weigelt}, G.,
  {Woillez}, J., \& {de Zeeuw}, P.~T. 2004, \nat, 429, 47

\bibitem[{{Keel}(1980)}]{keel80}
{Keel}, W.~C. 1980, \aj, 85, 198

\bibitem[{{Krolik}(1999)}]{krolik99}
{Krolik}, J.~H. 1999, {Active galactic nuclei : from the central black hole to
  the galactic environment} (Active galactic nuclei : from the central black
  hole to the galactic environment /Julian H.~Krolik.~Princeton, N.~J.~:
  Princeton University Press, c1999.)

\bibitem[La Franca et al.(2005)]{lafranca05} La Franca, F., et 
al.\ 2005, \apj, 635, 864 

\bibitem[{{Lacy} {et~al.}(2007){Lacy}, {Petric}, {Sajina}, {Canalizo},
  {Storrie-Lombardi}, {Armus}, {Fadda}, \& {Marleau}}]{lacy07}
{Lacy}, M., {Petric}, A.~O., {Sajina}, A., {Canalizo}, G., {Storrie-Lombardi},
  L.~J., {Armus}, L., {Fadda}, D., \& {Marleau}, F.~R. 2007, \aj, 133, 186

\bibitem[Lacy et al.(2004)]{lacy04} Lacy, M., et al.\ 2004, 
\apjs, 154, 166 

\bibitem[{{Lawrence}(1991)}]{lawrence91}
{Lawrence}, A. 1991, \mnras, 252, 586

\bibitem[Lutz {et~al.}(2004)]{lutz04}
   Lutz, D., Maiolino, R., Spoon, H.W.W. \& Moorhead, A.F.M. 2004, A\& A, 418, 465

\bibitem[{{Maiolino} {et~al.}(2007){Maiolino}, {Shemmer}, {Imanishi}, {Netzer},
  {Oliva}, {Lutz}, \& {Sturm}}]{maiolino07}
{Maiolino}, R., {Shemmer}, O., {Imanishi}, M., {Netzer}, H., {Oliva}, E.,
  {Lutz}, D., \& {Sturm}, E. 2007, \aap, 468, 979

\bibitem[{{Makovoz} \& {Marleau}(2005)}]{makovoz05}
{Makovoz}, D. \& {Marleau}, F.~R. 2005, \pasp, 117, 1113

\bibitem[{{Marconi} {et~al.}(2004){Marconi}, {Risaliti}, {Gilli}, {Hunt},
  {Maiolino}, \& {Salvati}}]{marconi04}
{Marconi}, A., {Risaliti}, G., {Gilli}, R., {Hunt}, L.~K., {Maiolino}, R., \&
  {Salvati}, M. 2004, \mnras, 351, 169

\bibitem[{{Martin} {et~al.}(2005)}]{martin05}
{Martin}, D.~C. {et~al.} 2005, \apjl, 619, L1

\bibitem[{{Mart{\'{\i}}nez-Sansigre} {et~al.}(2006){Mart{\'{\i}}nez-Sansigre},
  {Rawlings}, {Lacy}, {Fadda}, {Jarvis}, {Marleau}, {Simpson}, \&
  {Willott}}]{martinez06}
{Mart{\'{\i}}nez-Sansigre}, A., {Rawlings}, S., {Lacy}, M., {Fadda}, D.,
  {Jarvis}, M.~J., {Marleau}, F.~R., {Simpson}, C., \& {Willott}, C.~J. 2006,
  \mnras, 370, 1479

\bibitem[{{McLeod} \& {Rieke}(1995)}]{mcleod95}
{McLeod}, K.~K. \& {Rieke}, G.~H. 1995, \apj, 441, 96

\bibitem[{{Nenkova} {et~al.}(2002){Nenkova}, {Ivezi{\' c}}, \&
  {Elitzur}}]{nenkova02}
{Nenkova}, M., {Ivezi{\' c}}, {\v Z}., \& {Elitzur}, M. 2002, \apjl, 570, L9

\bibitem[{{Pier} \& {Krolik}(1992)}]{pier92}
{Pier}, E.~A. \& {Krolik}, J.~H. 1992, \apj, 401, 99

\bibitem[{{Pier} \& {Krolik}(1993)}]{pier93}
---. 1993, \apj, 418, 673

\bibitem[Polletta et al.(2006)]{polletta06} Polletta, M.~d.~C., et 
al.\ 2006, \apj, 642, 673 

\bibitem[{{Richards} {et~al.}(2006){Richards}, {Lacy}, {Storrie-Lombardi},
  {Hall}, {Gallagher}, {Hines}, {Fan}, {Papovich}, {Vanden Berk}, {Trammell},
  {Schneider}, {Vestergaard}, {York}, {Jester}, {Anderson}, {Budav{\'a}ri}, \&
  {Szalay}}]{richards06}
{Richards}, G.~T., {Lacy}, M., {Storrie-Lombardi}, L.~J., {Hall}, P.~B.,
  {Gallagher}, S.~C., {Hines}, D.~C., {Fan}, X., {Papovich}, C., {Vanden Berk},
  D.~E., {Trammell}, G.~B., {Schneider}, D.~P., {Vestergaard}, M., {York},
  D.~G., {Jester}, S., {Anderson}, S.~F., {Budav{\'a}ri}, T., \& {Szalay},
  A.~S. 2006, \apjs, 166, 470

\bibitem[{{Rigby} {et~al.}(2006){Rigby}, {Rieke}, {Donley}, {Alonso-Herrero},
  \& {P{\'e}rez-Gonz{\'a}lez}}]{rigby06}
{Rigby}, J.~R., {Rieke}, G.~H., {Donley}, J.~L., {Alonso-Herrero}, A., \&
  {P{\'e}rez-Gonz{\'a}lez}, P.~G. 2006, \apj, 645, 115

\bibitem[{{Risaliti} {et~al.}(1999){Risaliti}, {Maiolino}, \&
  {Salvati}}]{risaliti99}
{Risaliti}, G., {Maiolino}, R., \& {Salvati}, M. 1999, \apj, 522, 157

\bibitem[Sanders et al.(2007)]{sanders07} Sanders, D.~B., et al.\ 
2007, \apjs, 172, 86 

\bibitem[{{Schmitt} {et~al.}(2003){Schmitt}, {Donley}, {Antonucci},
  {Hutchings}, \& {Kinney}}]{schmitt03}
{Schmitt}, H.~R., {Donley}, J.~L., {Antonucci}, R.~R.~J., {Hutchings}, J.~B.,
  \& {Kinney}, A.~L. 2003, \apjs, 148, 327

\bibitem[Schweitzer et al.(2006)]{schweitzer06} Schweitzer, M., et 
al.\ 2006, \apj, 649, 79 

\bibitem[Shi et al.(2007)]{shi07} Shi, Y., et al.\ 2007, 
\apj, 669, 841 

\bibitem[{{Simpson}(2005)}]{simpson05}
{Simpson}, C. 2005, \mnras, 360, 565

\bibitem[{{Spergel} {et~al.}(2007){Spergel}, {Bean}, {Dor{\'e}}, {Nolta},
  {Bennett}, {Dunkley}, {Hinshaw}, {Jarosik}, {Komatsu}, {Page}, {Peiris},
  {Verde}, {Halpern}, {Hill}, {Kogut}, {Limon}, {Meyer}, {Odegard}, {Tucker},
  {Weiland}, {Wollack}, \& {Wright}}]{spergel07}
{Spergel}, D.~N., {Bean}, R., {Dor{\'e}}, O., {Nolta}, M.~R., {Bennett}, C.~L.,
  {Dunkley}, J., {Hinshaw}, G., {Jarosik}, N., {Komatsu}, E., {Page}, L.,
  {Peiris}, H.~V., {Verde}, L., {Halpern}, M., {Hill}, R.~S., {Kogut}, A.,
  {Limon}, M., {Meyer}, S.~S., {Odegard}, N., {Tucker}, G.~S., {Weiland},
  J.~L., {Wollack}, E., \& {Wright}, E.~L. 2007, \apjs, 170, 377

\bibitem[{{Steffen} {et~al.}(2003){Steffen}, {Barger}, {Cowie}, {Mushotzky}, \&
  {Yang}}]{steffen03}
{Steffen}, A.~T., {Barger}, A.~J., {Cowie}, L.~L., {Mushotzky}, R.~F., \&
  {Yang}, Y. 2003, \apjl, 596, L23

\bibitem[{{Stern} {et~al.}(2005){Stern}, {Eisenhardt}, {Gorjian}, {Kochanek},
  {Caldwell}, {Eisenstein}, {Brodwin}, {Brown}, {Cool}, {Dey}, {Green},
  {Jannuzi}, {Murray}, {Pahre}, \& {Willner}}]{stern05}
{Stern}, D., {Eisenhardt}, P., {Gorjian}, V., {Kochanek}, C.~S., {Caldwell},
  N., {Eisenstein}, D., {Brodwin}, M., {Brown}, M.~J.~I., {Cool}, R., {Dey},
  A., {Green}, P., {Jannuzi}, B.~T., {Murray}, S.~S., {Pahre}, M.~A., \&
  {Willner}, S.~P. 2005, \apj, 631, 163

\bibitem[{{Telfer} {et~al.}(2002){Telfer}, {Zheng}, {Kriss}, \&
  {Davidsen}}]{telfer02}
{Telfer}, R.~C., {Zheng}, W., {Kriss}, G.~A., \& {Davidsen}, A.~F. 2002, \apj,
  565, 773

\bibitem[{{Trammell} {et~al.}(2007){Trammell}, {Vanden Berk}, {Schneider},
  {Richards}, {Hall}, {Anderson}, \& {Brinkmann}}]{trammell07}
{Trammell}, G.~B., {Vanden Berk}, D.~E., {Schneider}, D.~P., {Richards}, G.~T.,
  {Hall}, P.~B., {Anderson}, S.~F., \& {Brinkmann}, J. 2007, \aj, 133, 1780

\bibitem[{{Treister} \& {Urry}(2006)}]{treister06b}
{Treister}, E. \& {Urry}, C.~M. 2006, \apjl, 652, L79

\bibitem[{{Treister} {et~al.}(2006){Treister}, {Urry}, {Van Duyne},
  {Dickinson}, {Chary}, {Alexander}, {Bauer}, {Natarajan}, {Lira}, \&
  {Grogin}}]{treister06a}
{Treister}, E., {Urry}, C.~M., {Van Duyne}, J., {Dickinson}, M., {Chary},
  R.-R., {Alexander}, D.~M., {Bauer}, F., {Natarajan}, P., {Lira}, P., \&
  {Grogin}, N.~A. 2006, \apj, 640, 603

\bibitem[{{Treister} {et~al.}(2004)}]{treister04}
{Treister}, E. {et~al.} 2004, \apj, 616, 123

\bibitem[{{Tristram} {et~al.}(2007){Tristram}, {Meisenheimer}, {Jaffe},
  {Schartmann}, {Rix}, {Leinert}, {Morel}, {Wittkowski}, {R{\"o}ttgering},
  {Perrin}, {Lopez}, {Raban}, {Cotton}, {Graser}, {Paresce}, \&
  {Henning}}]{tristram07}
{Tristram}, K.~R.~W., {Meisenheimer}, K., {Jaffe}, W., {Schartmann}, M., {Rix},
  H.~., {Leinert}, C., {Morel}, S., {Wittkowski}, M., {R{\"o}ttgering}, H.,
  {Perrin}, G., {Lopez}, B., {Raban}, D., {Cotton}, W.~D., {Graser}, U.,
  {Paresce}, F., \& {Henning}, T. 2007, A\&A in press, arXiv:0709.0209

\bibitem[{{Trump} {et~al.}(2006){Trump}, {Impey}, {McCarthy}, {Elvis},
  {Huchra}, {Brusa}, {Hasinger}, {Schinnerer}, {Capak}, {Lilly}, \&
  {Scoville}}]{trump06}
{Trump}, J.~R., {Impey}, C.~D., {McCarthy}, P.~J., {Elvis}, M., {Huchra},
  J.~P., {Brusa}, M., {Hasinger}, G., {Schinnerer}, E., {Capak}, P., {Lilly},
  S.~J., \& {Scoville}, N.~Z. 2006, ArXiv Astrophysics e-prints

\bibitem[{{Ueda} {et~al.}(2003){Ueda}, {Akiyama}, {Ohta}, \& {Miyaji}}]{ueda03}
{Ueda}, Y., {Akiyama}, M., {Ohta}, K., \& {Miyaji}, T. 2003, \apj, 598, 886

\bibitem[{{van Bemmel} \& {Dullemond}(2003)}]{vanbemmel03}
{van Bemmel}, I.~M. \& {Dullemond}, C.~P. 2003, \aap, 404, 1

\bibitem[{{Vanden Berk} {et~al.}(2004){Vanden Berk}, {Wilhite}, {Kron},
  {Anderson}, {Brunner}, {Hall}, {Ivezi{\'c}}, {Richards}, {Schneider}, {York},
  {Brinkmann}, {Lamb}, {Nichol}, \& {Schlegel}}]{vandenberk04}
{Vanden Berk}, D.~E., {Wilhite}, B.~C., {Kron}, R.~G., {Anderson}, S.~F.,
  {Brunner}, R.~J., {Hall}, P.~B., {Ivezi{\'c}}, {\v Z}., {Richards}, G.~T.,
  {Schneider}, D.~P., {York}, D.~G., {Brinkmann}, J.~V., {Lamb}, D.~Q.,
  {Nichol}, R.~C., \& {Schlegel}, D.~J. 2004, \apj, 601, 692

\bibitem[{{Yip} {et~al.}(2004){Yip}, {Connolly}, {Vanden Berk}, {Ma},
  {Frieman}, {SubbaRao}, {Szalay}, {Richards}, {Hall}, {Schneider}, {Hopkins},
  {Trump}, \& {Brinkmann}}]{yip04}
{Yip}, C.~W., {Connolly}, A.~J., {Vanden Berk}, D.~E., {Ma}, Z., {Frieman},
  J.~A., {SubbaRao}, M., {Szalay}, A.~S., {Richards}, G.~T., {Hall}, P.~B.,
  {Schneider}, D.~P., {Hopkins}, A.~M., {Trump}, J., \& {Brinkmann}, J. 2004,
  \aj, 128, 2603

\bibitem[{{York} {et~al.}(2000)}]{york00}
{York}, D.~G. {et~al.} 2000, \aj, 120, 1579

\bibitem[{{Zakamska} {et~al.}(2003){Zakamska}, {Strauss}, {Krolik}, {Collinge},
  {Hall}, {Hao}, {Heckman}, {Ivezi{\'c}}, {Richards}, {Schlegel}, {Schneider},
  {Strateva}, {Vanden Berk}, {Anderson}, \& {Brinkmann}}]{zakamska03}
{Zakamska}, N.~L., {Strauss}, M.~A., {Krolik}, J.~H., {Collinge}, M.~J.,
  {Hall}, P.~B., {Hao}, L., {Heckman}, T.~M., {Ivezi{\'c}}, {\v Z}.,
  {Richards}, G.~T., {Schlegel}, D.~J., {Schneider}, D.~P., {Strateva}, I.,
  {Vanden Berk}, D.~E., {Anderson}, S.~F., \& {Brinkmann}, J. 2003, \aj, 126,
  2125

\end{thebibliography}
\end{document}